\documentclass[10pt,conference]{IEEEtran}
\IEEEoverridecommandlockouts
\usepackage{cite}
\usepackage{amsmath,amssymb,amsfonts}
\usepackage{algorithmic}
\usepackage{graphicx}
\usepackage{textcomp}
\usepackage{listings}
\usepackage{makecell}
\usepackage{xcolor}
\usepackage{xcolor}
\usepackage{xurl}
\usepackage[hidelinks]{hyperref}
\usepackage{tcolorbox}
\usepackage{xcolor}
\usepackage{listings}
\usepackage{caption}

\definecolor{answer-gray}{gray}{0.92}  
\newtcolorbox{answerbox}{
    colback=answer-gray,    
    colframe=answer-gray,   
    boxrule=0pt,
    arc=2pt,                
    left=6pt,
    right=6pt,
    top=4pt,
    bottom=4pt,
    boxsep=0pt
}
\def\BibTeX{{\rm B\kern-.05em{\sc i\kern-.025em b}\kern-.08em
    T\kern-.1667em\lower.7ex\hbox{E}\kern-.125emX}}
\usepackage{booktabs}

\newcommand{\ym}[1]{\textcolor{black}{{#1}}}
\usepackage{xspace}
\newcommand{\tool}{Memoir\xspace}
\def\BibTeX{{\rm B\kern-.05em{\sc i\kern-.025em b}\kern-.08em
    T\kern-.1667em\lower.7ex\hbox{E}\kern-.125emX}}
\begin{document}

\title{Memoir: Learning, Verifying, and Evolving False-Positive Memories for Static Application Security Testing Tools\\
}

\author{

\centering

\makebox[\textwidth][c]{
\parbox{0.88\textwidth}{

\centering

{\large
Shenyuan Guan\textsuperscript{1},
Qiaodan Hou\textsuperscript{2},
Yanjun Chen\textsuperscript{2},
Xincheng Wen\textsuperscript{1},
Jia Feng\textsuperscript{1},
Keke Lian\textsuperscript{2},
Cuiyun Gao\textsuperscript{1*}
\par}

\vspace{0.6em}


\vspace{0.6em}


}

}

}

\maketitle


\begin{abstract}

Static Application Security Testing (SAST) tools have become 
indispensable in modern secure software development. However, these tools often generate 
false-positive (FP) alerts, imposing substantial manual inspection costs and 
\ym{reducing the trust from developers}. Existing FP reduction 
\ym{methods} still face two primary challenges. \ym{First, the large differences among SAST tools and vulnerability categories make it difficult for these methods to learn recurring patterns in historical false positives. Moreover, the knowledge used by these methods are largely static and cannot be updated as newly validated cases accumulate.} 

To address these challenges, we propose \textit{\tool}, a memory-driven 
framework 
\ym{for identifying false positives by transforming historical FP alerts into reusable semantic memories.} It consists of two key modules. \ym{First,} 
\emph{historical semantic memory construction} 
 \ym{converts} historical FP alerts into structured semantic memories through LLM-guided annotation, pattern clustering, and memory synthesis to capture reusable behavioral patterns. Moreover, \emph{memory-driven identification and evolution} retrieves relevant memories and performs semantic verification against taxonomy consistency and security invariants before making the final prediction. 
 \ym{It then} incorporates verified predictions back into the memory repository, \ym{allowing the knowledge base to evolve as new cases accumulate}. 
We evaluate \tool on CWE-Bench-Java to demonstrate its effectiveness in real-world security analysis.
Specifically, with GPT-4o as the backbone, \tool achieves an FP-F1 of 98.82\%, with an FP-Rec of 100.00\% and an FP-Precision of 97.66\%, consistently outperforming all compared baselines in false-positive identification.
Furthermore, an industrial case study on production software systems from Tencent shows that the learned memory base generalizes effectively across different SAST tools without retraining.
\end{abstract}


\section{Introduction}
Static Application Security Testing (SAST) has become an indispensable component of modern secure software development for detecting vulnerabilities such as SQL injection (CWE-089), path traversal (CWE-022), and command injection (CWE-078) without executing programs. Industrial SAST tools, including CodeQL~\cite{codeql}, Checkmarx~\cite{checkmarx}, and Fortify~\cite{fortify}, are routinely integrated into continuous integration and continuous deployment (CI/CD) pipelines to identify security flaws before software release. By statically analyzing source code, these tools help developers discover vulnerabilities early and reduce the risk of deploying insecure software.

Despite their widespread adoption, the practical effectiveness of SAST tools is severely limited by the large number of false-positive alerts. Previous studies have reported that false positives frequently account for more than half of all reported warnings in real-world software projects~\cite{johnson2013dont, christakis2016developers}. On the public CWE-Bench-Java benchmark~\cite{cwebenchjava}, our evaluation shows that CodeQL reports 865 alerts across 120 Java projects, of which \textbf{95.66\%} are manually verified as false positives. Such overwhelming numbers of spurious alerts substantially increase manual triage effort, create alert fatigue, and ultimately reduce developers' trust in static analysis tools~\cite{bessey2010coverity}.

Existing approaches for false-positive reduction generally fall into three categories~\cite{heckman2011actionable, muske2016survey}. \emph{Rule-based} techniques manually encode expert knowledge to suppress known false-positive patterns~\cite{kremenek2004zranking, sadowski2015tricorder}; while effective in specific scenarios, these handcrafted rules require continuous maintenance and are difficult to transfer across programming languages, analysis tools, or software frameworks. \emph{Learning-based} approaches formulate false-positive identification as a supervised classification problem; although they achieve encouraging results on benchmark datasets, they heavily depend on labeled training data and often fail to generalize to unseen projects or vulnerability categories. More recently, Large Language Models (LLMs) have demonstrated remarkable reasoning capabilities for software engineering tasks, and \emph{retrieval-based reasoning} methods typically rely on zero-shot prompting or Retrieval-Augmented Generation (RAG) to analyze each alert independently. Despite their improvements in specific settings, these approaches still face two primary challenges:

\lstdefinestyle{fpListing}{
  language=Java,
  basicstyle=\ttfamily\scriptsize,
  numbers=left,
  numberstyle=\tiny,
  stepnumber=1,
  numbersep=6pt,
  frame=single,
  columns=fullflexible,
  breaklines=true,
  keepspaces=true,
  showstringspaces=false,
  captionpos=t,
  keywordstyle=\bfseries,
  commentstyle=\itshape\color{green!45!black},
  stringstyle=\color{purple},
  xleftmargin=1.5em,
  framexleftmargin=1.2em,
  aboveskip=0.4em,
  belowskip=0.4em
}

\textbf{(1) Difficulty in learning complex FP semantics.}
False-positive alerts are not isolated events but \emph{recurring semantic phenomena}: although they may be reported from different repositories, frameworks, and vulnerability categories, many of them share common semantic characteristics. As illustrated in Listing~\ref{lst:recurring_fp_examples}, whitelist mapping, framework-managed escaping, prepared-statement binding, and path normalization with boundary checks repeatedly explain why tainted input cannot control a security-sensitive operation, even across entirely different CWE categories. On CWE-Bench-Java, a single sanitization pattern (e.g., framework-managed escaping in Spring/Thymeleaf) alone accounts for the false-positive rationale of over 30\% of XSS alerts across more than 15 distinct projects. Therefore, these cases forced developers to repeatedly rediscover the same evidence, severely limiting their generalization across projects, analysis tools, and vulnerability categories.

\textbf{(2) Lack of continuously updatable knowledge.}
Even when recurring rationales have been captured, the knowledge base that stores them must keep pace with an ever-changing software landscape: codebases are refactored, frameworks are upgraded, and new vulnerability categories emerge on a regular basis. However, existing FP reduction techniques rely on largely static knowledge that cannot be incrementally updated as new cases accumulate. Prior experience progressively drifts out of alignment with the evolving codebase and vulnerability landscape, causing the effectiveness of these techniques to gradually deteriorate over time.

\begin{lstlisting}[
style=fpListing,
caption={Examples of recurring false-positive semantics across different vulnerability types.},
label={lst:recurring_fp_examples}
]
/* (a) Path Traversal: Whitelist mapping */
String name = req.getParameter("file");

String safe = switch (name) {
  case "report" -> "report.pdf";
  case "log"    -> "server.log";
  default       -> "default.txt";
};

Path p = BASE.resolve(safe);
return Files.readString(p);


/* (b) XSS: Framework-managed escaping */
String msg = req.getParameter("msg");

model.addAttribute("msg", msg);
return "result";

// result.html
// <span th:text="${msg}"></span>


/* (c) Command Injection: Constant command selection */
String act = req.getParameter("action");

String cmd = switch (act) {
  case "status" -> "/bin/status";
  case "clean"  -> "/bin/clean";
  default       -> "/bin/help";
};

new ProcessBuilder(cmd).start();


/* (d) SQL Injection: Parameter binding */
String id = req.getParameter("id");

PreparedStatement ps =
  conn.prepareStatement(
    "SELECT * FROM users WHERE id=?");

ps.setString(1, id);
ResultSet rs = ps.executeQuery();


/* (e) Code Injection: Type whitelist */
String type = req.getParameter("type");

Class<?> c = switch (type) {
  case "json" -> JsonParser.class;
  case "xml"  -> XmlParser.class;
  default     -> SafeParser.class;
};

Parser p = (Parser)
  c.getConstructor().newInstance();


/* (f) Path Traversal: Security invariant */
String file = req.getParameter("file");

Path p = BASE.resolve(file).normalize();

if (!p.startsWith(BASE)) {
  throw new SecurityException();
}

return Files.readString(p);
\end{lstlisting}

\textbf{Our work.}
To address these challenges, we propose \textit{\tool}, a \textbf{memory-driven false-positive identification framework} that reformulates SAST false-positive identification as a reusable semantic memory learning problem. Rather than treating each alert as an independent reasoning task, \tool learns, verifies, and continuously evolves reusable false-positive memories, so that recurring rationales are abstracted \emph{once} and then reused across projects, tools, and vulnerability categories.

Specifically, \textit{\tool} consists of two key modules that directly target the two challenges above:
(1) A \emph{historical semantic memory construction} module, which distills historical false-positive alerts into structured semantic memories through LLM-guided annotation, semantic clustering, and memory synthesis. Each memory captures reusable security knowledge, including false-positive taxonomy, security invariants, taint-breaking semantics, and representative evidence, thereby turning heterogeneous historical alerts into transferable knowledge units; and
(2) A \emph{memory-driven identification and evolution} module, which, for each new SAST alert, retrieves the most relevant memories and performs semantic verification against taxonomy consistency and security invariants before producing the final prediction. Verified cases are subsequently incorporated back into the memory repository, enabling the memory base to continuously evolve as new cases accumulate and to progressively improve future reasoning.

We evaluate \tool on the public \textbf{CWE-Bench-Java} benchmark~\cite{cwebenchjava}, using CodeQL-generated alerts. Compared with representative baselines including Zero-shot prompting, IRIS~\cite{li2025iris}, conventional Retrieval-Augmented Generation, BugLens~\cite{li2025buglens}, and ZeroFalse~\cite{iranmanesh2025zerofalse}, \tool consistently achieves the best overall performance, reaching an \textbf{FP-F1 of 98.82\%}, \textbf{FP-Recall of 100.00\%}, and \textbf{FP-Precision of 97.66\%}.
Furthermore, industrial validation on large-scale production software from \textbf{Tencent} demonstrates that the learned memory repository generalizes effectively across different SAST tools without retraining, reducing false positives by \textbf{80.0\%} for CodeQL and \textbf{94.6\%} for XCheck. These results demonstrate that reusable and continuously evolving false-positive memories provide an effective and generalizable foundation for improving the practicality of modern SAST systems.

\textbf{Contributions.} The major contributions of this paper are summarized as follows:
\begin{enumerate}
    \item We propose \textit{\tool}, a memory-driven framework that reformulates SAST false-positive identification as a reusable semantic memory learning problem, jointly addressing the difficulty of learning complex FP semantics and the lack of continuously updatable knowledge in existing techniques.

    \item We design a structured memory architecture together with a historical semantic memory construction module and a memory-driven identification and evolution module, allowing historical diagnosis experience to be abstracted, verified, and continuously reused across projects, analysis tools, and vulnerability categories.

    \item We evaluate \textit{\tool} on the public \textbf{CWE-Bench-Java} benchmark and on large-scale industrial software projects from Tencent. Results show that \tool consistently outperforms representative LLM-based baselines and substantially improves the practicality of modern SAST tools.
\end{enumerate}

\section{Proposed Framework}
\label{sec:framework}

\subsection{Problem Formulation}
\label{subsec:problem_formulation}

We formulate SAST false-positive identification as a memory-guided decision problem. 
Unlike conventional alert classification approaches that independently analyze each warning, Memoir aims to exploit historical triage experience by learning reusable false-positive patterns from previously verified alerts.

Existing SAST false-positive reduction approaches mainly focus on alert-level reasoning, where each warning is analyzed independently based on its local context. However, many false positives are caused by recurring semantic patterns, such as framework-provided protections, implicit validation mechanisms, secure data transformations, or non-user-controlled data flows. These patterns are repeatedly observed during security triage but cannot be effectively reused by conventional approaches. Memoir addresses this limitation by explicitly modeling historical triage knowledge as structured false-positive memories.

A CodeQL alert is represented as:
\begin{equation}
A=\langle trace,cwe,rule,source,sink,analysis\rangle ,
\end{equation}
where $trace$ denotes the taint propagation path, $cwe$ and $rule$ represent vulnerability metadata, and $source$, $sink$, and $analysis$ describe the taint endpoints and analyzer information. Each alert has a ground-truth label $y\in\{FP,TP\}$.

Given a set of historical false-positive alerts, Memoir constructs a false-positive memory repository:
\begin{equation}
\mathcal{M}=\{m_1,m_2,\dots,m_n\},
\end{equation}
where each memory entry encodes a generalized false-positive pattern with semantic security knowledge rather than a single alert instance. The objective is to learn a decision function:
\begin{equation}
f(A,\mathcal{M})\rightarrow\{FP,TP\},
\end{equation}
which identifies false positives while preserving true vulnerabilities.

\subsection{Framework Overview}
\label{subsec:overview}

\begin{figure*}[t]
    \centering
    \includegraphics[width=0.85\linewidth]{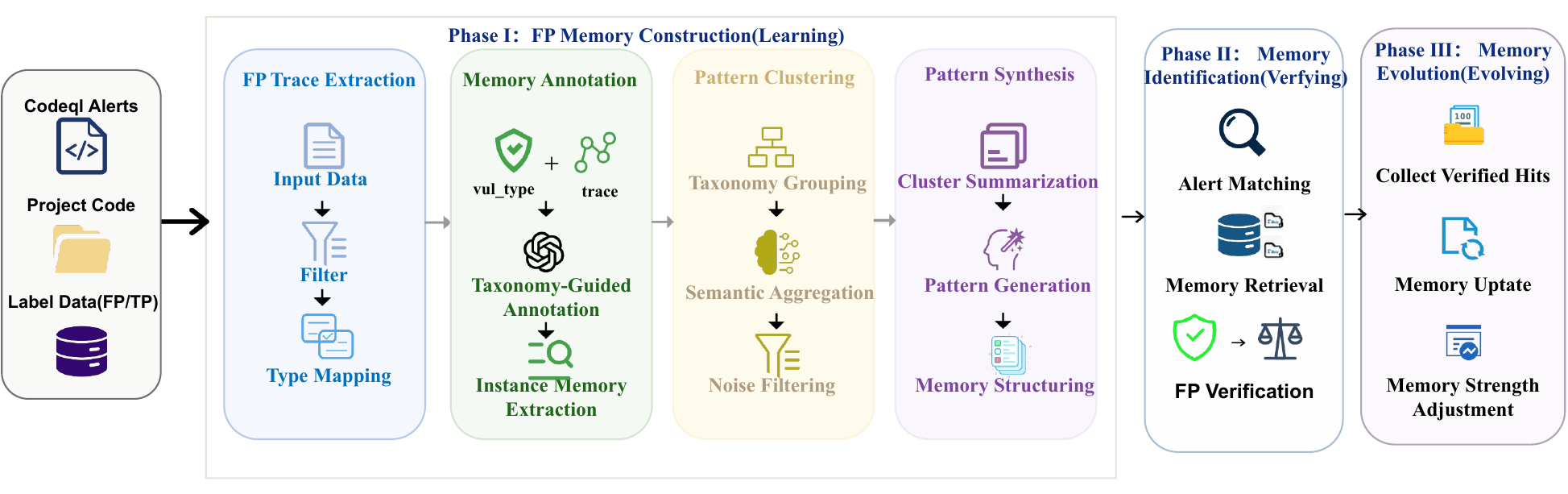}
    \caption{Overall architecture of Memoir.}
    \label{fig:overall_arch}
\end{figure*}

Figure~\ref{fig:overall_arch} presents the overall architecture of Memoir, a learning-verifying-evolving framework for SAST false-positive identification. 
The key idea is to transform scattered historical triage results into structured memories that can be retrieved, verified, and continuously refined.

Memoir follows three stages: 
(1) constructing reusable false-positive memories from historical alerts, 
(2) identifying new alerts through memory retrieval and semantic verification, and 
(3) evolving memories using verified feedback.

Different from conventional retrieval-based methods that directly reuse historical examples, Memoir performs knowledge abstraction before reuse. Instead of memorizing individual alerts, Memoir summarizes multiple similar cases into generalized patterns with explicit security reasoning elements. Therefore, retrieved memories provide not only similar examples but also interpretable evidence for final decision making.

The memory repository is organized into three complementary levels.

\textbf{Taxonomy Memory} stores coarse-grained false-positive categories, CWE information, and semantic tags for efficient memory filtering.

\textbf{Pattern Memory} captures reusable false-positive knowledge, including security invariants, taint breakers, retrieval keys, verification rules, and representative examples.

\textbf{Evolution Memory} records historical verification feedback and usage statistics, allowing the memory repository to adapt to newly observed code patterns.

\subsection{False-Positive Memory Construction}
\label{subsec:memory_construction}

The goal of this stage is to transform historical case-level false positives into reusable pattern-level memories.

Each memory entry is represented as:
\begin{equation}
m=\langle T,I,B,K,V,E\rangle ,
\end{equation}
where $T$, $I$, $B$, $K$, $V$, and $E$ denote taxonomy information, security invariants, taint breakers, retrieval keys, verification rules, and representative examples, respectively.

Given historical CodeQL alerts and triage labels, Memoir first extracts standardized false-positive traces containing taint paths, CWE information, source/sink contexts, and relevant code semantics.

Instead of directly storing raw alerts, Memoir applies LLM-based semantic annotation to identify the underlying reasons why an alert is considered a false positive, including false-positive categories, root causes, security invariants, and taint-breaking mechanisms.

The annotated cases are then clustered according to semantic similarity over invariants, taint breakers, and security-related features. Unlike surface-level similarity clustering, Memoir focuses on semantic factors that determine whether an alert is actually a false positive, such as whether input is constrained by validation, transformed by secure APIs, or protected by framework-level mechanisms.

Cases belonging to the same cluster are synthesized into a generalized memory entry that summarizes the shared false-positive pattern and provides retrieval keys and verification rules. The synthesized memory therefore represents reusable security knowledge rather than a collection of historical examples.

Through this abstraction process, Memoir converts individual triage decisions into structured security knowledge, enabling knowledge transfer across different alerts, projects, and vulnerability instances.

\subsection{Memory-Guided False-Positive Identification}
\label{subsec:identification}

\begin{figure}[t]
    \centering
    \includegraphics[width=\linewidth]{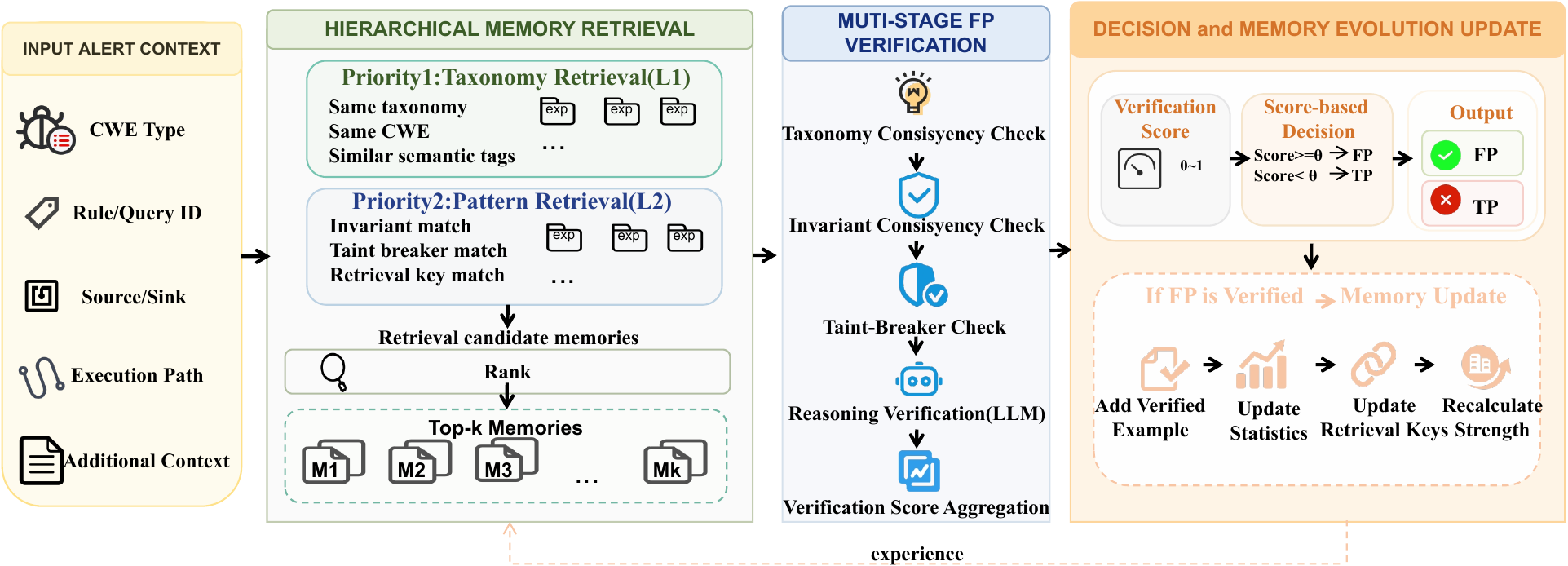}
    \caption{Memory retrieval and verification workflow.}
    \label{fig:workflow}
\end{figure}

Given a new CodeQL alert, Memoir retrieves relevant memories and verifies their semantic consistency before producing the final decision, as shown in Figure~\ref{fig:workflow}.

The main challenge during identification is that retrieved memories may be relevant at the lexical level but invalid at the security semantic level. For example, two alerts may share similar source and sink locations but differ in whether the data is user-controlled or protected by validation logic. Therefore, Memoir separates retrieval from verification instead of directly applying historical decisions.






Memoir adopts a hierarchical retrieval strategy to balance retrieval efficiency and semantic accuracy. It first performs taxonomy-level filtering based on CWE categories, rule types, and semantic tags to eliminate irrelevant memories. The remaining candidates are then ranked using multiple semantic signals, including security invariant similarity, taint-breaker consistency, and retrieval-key matching, and the top-$K$ candidates are selected as verification evidence. To account for the fact that similar alerts may correspond to different security conditions, Memoir applies progressive verification before incorporating retrieved knowledge. This process assesses taxonomy consistency to ensure that the retrieved memory belongs to the same false-positive category as the current alert, and validates the alert against the security invariant and taint-breaking mechanism described in the memory. Finally, an LLM performs semantic reasoning over the alert context, retrieved memory, and verification evidence.

Only memories that pass verification contribute to the final FP/TP prediction, reducing the risk of incorrectly suppressing true vulnerabilities. Retrieved memories serve as supporting evidence rather than direct decisions, allowing Memoir to benefit from previous triage experience while maintaining conservative security judgment.

\subsection{Memory Evolution}
\label{subsec:evolution}

Static false-positive knowledge may become incomplete as software frameworks, coding practices, and security rules evolve. Memoir therefore introduces a closed-loop evolution mechanism that updates memories using verified results.

For each confirmed false-positive case, Memoir augments the corresponding memory with new examples, expands retrieval keys, and updates reliability statistics according to historical retrieval and verification outcomes.

These feedback signals allow frequently validated memories to become stronger evidence sources, while ineffective memories are gradually weakened through temporal decay.

Moreover, evolution enables Memoir to adapt to changes in software ecosystems, including newly introduced frameworks, coding conventions, and security practices.

Through continuous learning from verified decisions, Memoir evolves from a static knowledge repository into an adaptive false-positive memory system that improves over time.
\section{Experimental Setup}
\label{sec:setup}

This section describes the experimental setup used to evaluate Memoir. We first introduce the research questions, followed by the compared baselines, datasets, evaluation metrics, and implementation details.





\subsection{Research Questions}

We evaluate Memoir from three complementary perspectives.

\textbf{RQ1 (Overall Performance).} How effective is Memoir in identifying false-positive SAST alerts compared with existing LLM-based approaches, and does it maintain strong performance across different backbone models?

\textbf{RQ2 (Ablation Study).} How do Memoir's components and memory evolution mechanisms contribute to its performance and reliability?

\textbf{RQ3 (Industrial Applicability).} Can Memoir effectively reduce false-positive alerts and improve the practical usability of SAST tools in real-world industrial environments?

\subsection{Baselines}

We compare Memoir with four representative LLM-based false-positive identification approaches.

\textbf{Zero-shot} directly prompts the LLM to determine whether a CodeQL alert is a true vulnerability or a false positive using only the original alert information, without retrieval or external knowledge.

\textbf{IRIS}~\cite{li2025iris} is a recent LLM-based static analysis framework that performs structured reasoning over CodeQL alerts. Following its official setting, IRIS~\cite{li2025iris} receives the warning message, source, sink, and taint propagation path, and predicts whether the alert corresponds to a true vulnerability or a false positive.

\textbf{BugLens (adapted)}~\cite{li2025buglens} is a recent LLM-based post-refinement framework for taint-style static-analysis warnings. Following its structured analysis workflow, BugLens performs security-impact assessment and constraint analysis over each warning, and predicts whether the warning corresponds to a real vulnerability or a false positive. Since the original BugLens targets Linux kernel taint-style bugs, we adapt its reasoning workflow to our Java CodeQL alert setting by providing the warning message, source, sink, taint propagation path, vulnerability type, and relevant code context.

\textbf{Retrieval-Augmented Generation (RAG)} retrieves semantically similar historical false-positive cases and incorporates them into the LLM prompt as in-context examples. Unlike Memoir, conventional RAG does not construct abstract false-positive memories, does not organize them into reusable patterns, and does not perform explicit semantic verification after retrieval.

\textbf{ZeroFalse}~\cite{iranmanesh2025zerofalse} is a recent LLM-based false-positive reduction framework for static analysis. It improves the input evidence for LLM adjudication by enriching static-analysis reports with flow-sensitive traces, contextual evidence, and CWE-specific knowledge. Unlike Memoir, ZeroFalse~\cite{iranmanesh2025zerofalse}focuses on single-alert adjudication and does not explicitly learn reusable, verifiable, and evolvable false-positive memories from historical alerts.

For fair comparison, all approaches are evaluated using the same data split, alert representation, backbone model whenever applicable, and evaluation scripts.

\subsection{Public Benchmark Dataset}

For public evaluation, we use \textbf{CWE-Bench-Java}~\cite{cwebenchjava}, a manually verified benchmark proposed for evaluating LLM-assisted static application security testing on Java applications. We execute CodeQL on all repositories included in the benchmark and collect a total of 865 security alerts covering multiple CWE categories. Each alert is manually annotated as either a true positive (TP) or a false positive (FP) according to the benchmark ground truth.

To avoid data leakage, we divide the 865 alerts into training, validation, and test sets using a 7:1.5:1.5 ratio. The training set is exclusively used to construct the false-positive memory knowledge base, including false-positive trace extraction, taxonomy annotation, pattern clustering, memory synthesis, and memory evolution. The validation set is used for parameter tuning, such as retrieval thresholds and retrieval depth. The test set is reserved exclusively for final evaluation. No validation or test samples are used during memory construction.We ensure that the test projects are unseen during training and validation to evaluate cross-project generalization.

\subsection{Industrial Case Study Dataset}

In addition to the public benchmark, we conduct an industrial case study using an internal dataset collected from production software systems at Tencent. The dataset contains alerts generated by two widely used SAST tools, CodeQL and XCheck. All alerts were manually reviewed and labeled by experienced security engineers.

This industrial dataset is used only for practical validation and is not used in the controlled benchmark evaluation for RQ1--RQ3. Due to confidentiality constraints, the dataset cannot be released publicly, and we report only aggregated results, including false-positive reduction and precision improvement.

\subsection{Evaluation Metrics}


Following previous work, we report FP-Precision, FP-Recall, and FP-F1 for false-positive identification. FP-Precision measures the proportion of alerts predicted as false positives that are correctly classified, FP-Recall measures the proportion of verified false positives successfully identified, and FP-F1 is their harmonic mean.

To assess true-vulnerability preservation and overall classification performance, we additionally report TP-Recall, TP-F1, Macro-F1, and the Matthews correlation coefficient (MCC). TP-Recall measures the proportion of true vulnerabilities correctly identified:
\[
R_{\mathrm{TP}}=\frac{TP}{TP+FN},
\]
where \(TP\) and \(FN\) denote correctly detected and missed true vulnerabilities, respectively. TP-F1 is the F1-score for the true-vulnerability class. Macro-F1 is the average F1-score across the FP and TP classes:
\[
\mathrm{Macro\text{-}F1}=\frac{F1_{\mathrm{FP}}+F1_{\mathrm{TP}}}{2}.
\]
MCC measures overall binary classification performance:
\[
\mathrm{MCC}=\frac{TP\cdot TN-FP\cdot FN}
{\sqrt{(TP+FP)(TP+FN)(TN+FP)(TN+FN)}},
\]
where \(TP\), \(TN\), \(FP\), and \(FN\) follow the conventional binary-classification definitions. MCC ranges from \(-1\) to \(1\), with \(1\) indicating perfect predictions and \(0\) indicating no correlation between predictions and ground truth.

For memory retrieval, we report Mean Reciprocal Rank (MRR):
\[
\mathrm{MRR}=\frac{1}{|Q|}\sum_{q\in Q}\frac{1}{\mathrm{rank}_q},
\]
where \(Q\) is the set of retrieval queries and \(\mathrm{rank}_q\) is the rank of the first relevant memory; queries without a relevant memory contribute zero. MRR is not applicable to baselines without memory ranking.

For the industrial case study, we report false-positive reduction and precision improvement. False-positive reduction is calculated as:
\[
\mathrm{FP\ Reduction}=
\frac{FP_{\mathrm{before}}-FP_{\mathrm{after}}}
{FP_{\mathrm{before}}}\times100\%.
\]
Precision improvement is reported as the absolute change in precision and, where appropriate, its relative percentage increase.
\subsection{Implementation Details}

Memoir is implemented in Python and consists of seven sequential stages: false-positive trace extraction, memory annotation, pattern clustering, memory synthesis, memory matching, memory verification, and memory evolution.

Unless otherwise specified, GPT-4o~\cite{openai2024gpt4o} is used as the default backbone LLM. To evaluate backbone generalizability, we further replace GPT-4o~\cite{openai2024gpt4o} with Claude-3.5-Sonnet~\cite{anthropic2024claude35sonnet}, Gemini-1.5-Pro~\cite{geminiteam2024gemini15}, and DeepSeek-Coder-V2-Instruct~\cite{deepseekai2024deepseekcoderv2} while keeping all memory construction, retrieval, and verification components unchanged.

The memory retrieval module combines keyword-based matching with semantic similarity retrieval. By default, Memoir retrieves the Top-5 most relevant memory entries for each alert. Retrieved memories are then validated according to taxonomy consistency, security invariant matching, and semantic similarity before producing the final prediction. During test-time evaluation, the evolved memory knowledge base generated from the training set remains fixed, and no information from the evaluation data is incorporated into the memory repository.

All experiments are conducted using identical prompts, retrieval configurations, and evaluation scripts to ensure fair comparison across different methods and backbone models.

\section{Experimental Results and Analysis}
\label{sec:experiments}

This section presents the experimental results of Memoir on the public benchmark. We first compare Memoir with existing LLM-based baselines (RQ1), then quantify the contribution of key components through ablation study (RQ2),and finally evaluate the industrial applicability of Memoir in real-world SAST scenarios (RQ3).

\subsection{RQ1: Overall False-Positive Identification Performance}
\label{subsec:rq1}
\textbf{RQ1.} How effective is Memoir in identifying false-positive SAST alerts compared with existing LLM-based baselines?

To answer this question, we compare Memoir with Zero-shot prompting, IRIS~\cite{li2025iris}, conventional RAG, BugLens (adapted)~\cite{li2025buglens}, and ZeroFalse~\cite{iranmanesh2025zerofalse} on the manually verified CodeQL alert dataset. Table~\ref{tab:overall_performance} summarizes the overall results.

\begin{table*}[htbp]
\centering
\caption{Overall false positive identification performance comparison}
\label{tab:overall_performance}
\begin{tabular}{llcccccccc}
\toprule
Approach & Backbone Model & FP-Precision & FP-Rec & FP-F1 & MRR
& TP-Rec & TP-F1 & Macro-F1 & MCC \\
\midrule
Zero-shot & GPT-4o~\cite{openai2024gpt4o}
& 0.8421 & 0.0945 & 0.1702 & \textendash
& 0.1667 & 0.1818 & 0.1760 & 0.0470 \\
IRIS~\cite{li2025iris} & GPT-4o~\cite{openai2024gpt4o}
& 0.8835 & 0.3937 & 0.5436 & \textendash
& 0.3333 & 0.3158 & 0.4297 & 0.2441 \\
RAG & GPT-4o~\cite{openai2024gpt4o}
& 0.9012 & 0.6648 & 0.7655 & \textendash
& 0.3333 & 0.3636 & 0.5646 & 0.3624 \\
ZeroFalse~\cite{iranmanesh2025zerofalse} & GPT-4o~\cite{openai2024gpt4o}
& 0.9416 & 0.8472 & 0.8919 & \textendash
& 0.4167 & 0.4546 & 0.6732 & 0.4803 \\
BugLens (adapted)~\cite{li2025buglens} & GPT-4o~\cite{openai2024gpt4o}
& 0.9527 & 0.8615 & 0.9044 & \textendash
& 0.4167 & 0.4783 & 0.6914 & 0.5020 \\
\midrule
Memoir (Ours) & Gemini-1.5-Pro~\cite{geminiteam2024gemini15}
& 0.9640 & 0.9623 & 0.9623 & 0.9064
& 0.3333 & 0.4520 & 0.7072 & 0.5110 \\
Memoir (Ours) & DeepSeek-Coder-V2-Instruct~\cite{deepseekai2024deepseekcoderv2}
& 0.9562 & 0.9685 & 0.9623 & 0.9142
& 0.3333 & 0.4300 & 0.6962 & 0.4910 \\
Memoir (Ours) & Claude-3.5-Sonnet~\cite{anthropic2024claude35sonnet}
& 0.9721 & 0.9685 & 0.9703 & 0.9217
& 0.4167 & 0.5333 & 0.7518 & 0.5892 \\
Memoir (Ours) & GPT-4o~\cite{openai2024gpt4o}
& 0.9766 & 1.0000 & 0.9882 & 0.9398
& 0.5000 & 0.6667 & 0.8274 & 0.7006 \\
\bottomrule
\end{tabular}
\end{table*}

As shown in Table~\ref{tab:overall_performance}, Memoir achieves the strongest overall performance among the compared approaches when using GPT-4o as its backbone. It obtains an FP-Precision of 0.9766 and an FP-Recall of 1.0000, resulting in an FP-F1 score of 0.9882. Meanwhile, it achieves a TP-Recall of 0.5000, a TP-F1 score of 0.6667, a Macro-F1 of 0.8274, and an MCC of 0.7006. These results indicate that Memoir identifies all false-positive alerts in the evaluated test set while correctly retaining three of the six true-vulnerability alerts.

Among the existing baselines, BugLens (adapted) achieves an FP-Precision of 0.9527, an FP-Recall of 0.8615, and an FP-F1 score of 0.9044. ZeroFalse obtains an FP-F1 score of 0.8919, followed by conventional RAG at 0.7655, IRIS at 0.5436, and zero-shot prompting at 0.1702. Compared with BugLens (adapted), Memoir improves FP-F1 by 0.0838, or 8.38 percentage points. This improvement is accompanied by higher TP-Recall and MCC in the reported results, suggesting a better balance between identifying false-positive alerts and preserving true vulnerabilities.

We further evaluate Memoir with different backbone models while keeping the memory construction, retrieval, and verification pipeline unchanged. As shown in Table~\ref{tab:overall_performance}, GPT-4o achieves the highest FP-F1 score (0.9882) and Macro-F1 (0.8274) among the reported Memoir configurations. Claude-3.5-Sonnet, Gemini-1.5-Pro, and DeepSeek-Coder-V2-Instruct achieve FP-F1 scores of 0.9703, 0.9623, and 0.9623, respectively, with Macro-F1 scores of 0.7518, 0.7072, and 0.6962. Their MCC values are 0.5892, 0.5110, and 0.4910, respectively.

Although GPT-4o performs best across these reported metrics, the other backbones also achieve strong FP-Recall, ranging from 0.9623 to 0.9685. This suggests that Memoir's structured memory and verification pipeline can support different LLM backbones, while the choice of backbone still affects the balance between false-positive identification and true-vulnerability preservation.


\begin{answerbox}
\textbf{RQ1 Answer:}
Memoir achieves the strongest reported false-positive identification performance with GPT-4o, obtaining an FP-F1 score of 0.9882, a Macro-F1 of 0.8274, and an MCC of 0.7006. Compared with the strongest baseline, BugLens (adapted), it improves FP-F1 by 8.38 percentage points. Its TP-Recall of 0.5000 also highlights that true-vulnerability preservation remains an important challenge. Results across Claude-3.5-Sonnet, Gemini-1.5-Pro, and DeepSeek-Coder-V2-Instruct further indicate that Memoir's effectiveness extends beyond a single backbone, although performance varies across models.
\end{answerbox}

\subsection{RQ2: Ablation Study}

\label{subsec:rq2}

\textbf{RQ2.} How do Memoir's components and memory evolution mechanisms contribute to its performance and reliability?

To investigate the contribution of each module, we conduct component ablation studies by removing verification, removing clustering, and varying the retrieval depth. We further investigate the effects of memory evolution by comparing static memory, unverified evolution, and verified evolution. The first study examines the contribution of individual components, while the second evaluates whether feedback-driven memory updates improve detection performance and whether verification mitigates the risks of harmful updates.


\subsubsection{Component Ablation Study}

Table~\ref{tab:ablation} reports the results of component ablation and retrieval-depth experiments. We evaluate the contribution of verification and pattern clustering, and examine how retrieval depth affects false-positive identification performance using FP-Precision, FP-Rec, and FP-F1.

\begin{table}[htbp]
\centering
\caption{Ablation study results of Memoir components}
\label{tab:ablation}
\begin{tabular}{lccc}
\toprule
Configuration & FP-Precision & FP-Rec & FP-F1 \\
\midrule
Full Model & 0.9766 & 1.0000 & 0.9882 \\
\quad -- Verification & 0.9242 & 0.9720 & 0.9475 \\
\quad -- Clustering & 0.9510 & 0.8960 & 0.9227 \\
\quad Top-1 Retrieval & 0.9740 & 0.9310 & 0.9521 \\
\quad Top-3 Retrieval & 0.9752 & 0.9680 & 0.9716 \\
\quad Top-5 Retrieval & 0.9766 & 1.0000 & 0.9882 \\
\bottomrule
\end{tabular}
\end{table}






\subsubsection{Effect of Verification}

Removing the verification module reduces FP-Precision from 0.9766 to 0.9242, FP-Rec from 1.0000 to 0.9720, and FP-F1 from 0.9882 to 0.9475. The decline across all three metrics indicates that verification contributes to both the reliability and coverage of false-positive identification. Without verification, retrieved memories may be incorrectly applied to the current alert, leading to less reliable predictions. These results demonstrate the importance of validating the applicability of retrieved evidence before making a final classification.

\subsubsection{Effect of Pattern Clustering}

Removing pattern clustering reduces FP-Precision to 0.9510, FP-Rec to 0.8960, and FP-F1 to 0.9227. The decrease in all three metrics suggests that organizing historical false-positive cases into reusable patterns is beneficial for identifying new alerts. Without clustering, memories remain less effectively structured, potentially making retrieval more dependent on instance-level similarity and less capable of capturing recurring false-positive patterns. This result supports the role of pattern clustering in improving the reuse of historical false-positive knowledge.

\subsubsection{Effect of Retrieval Depth}

\begin{figure}[htbp]
\centering
\includegraphics[width=0.86\linewidth]{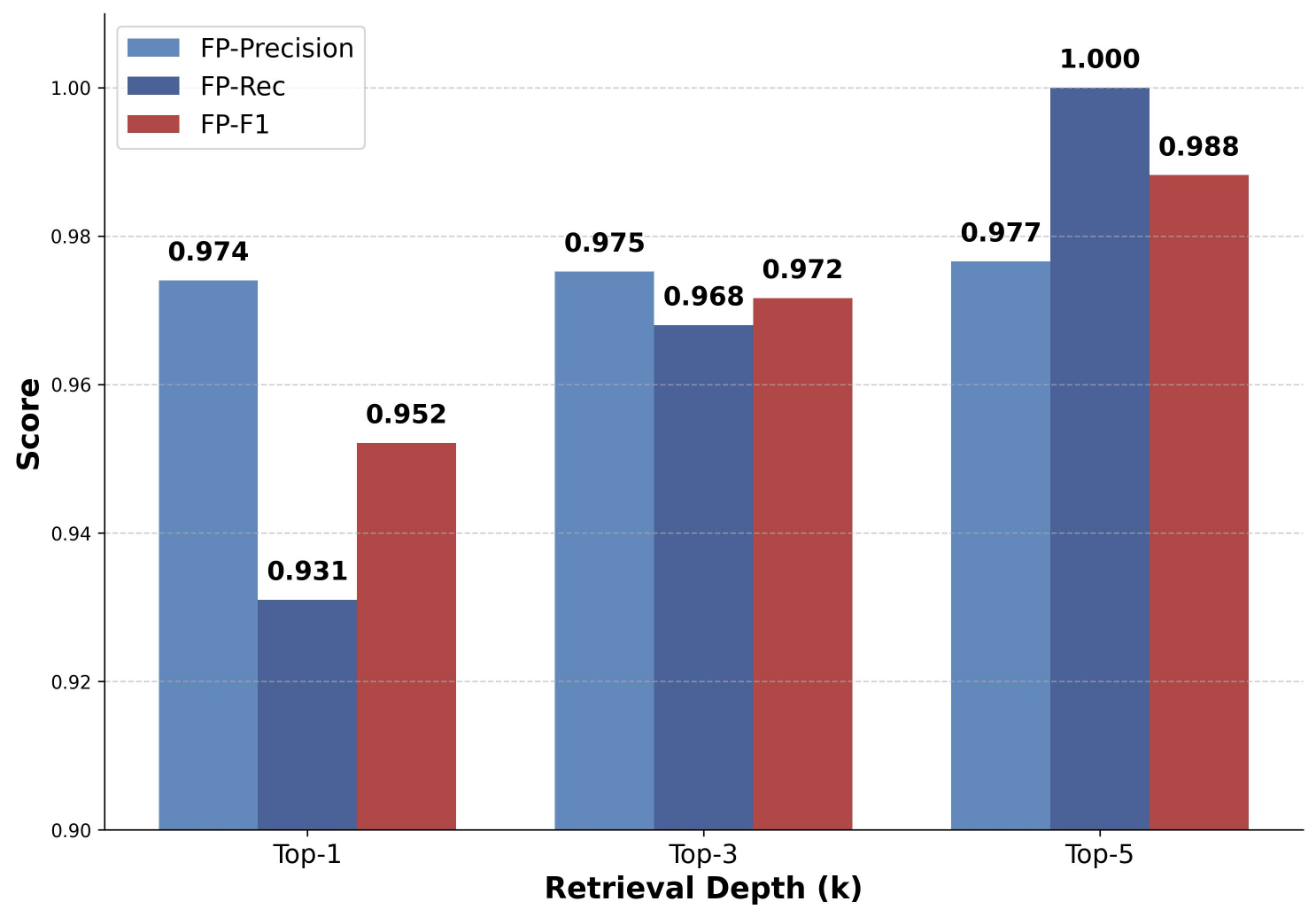}
\caption{Impact of retrieval depth on Memoir performance. Precision remains 1.000 across all Top-$k$ settings and is omitted for clarity. Increasing retrieval depth improves both Recall and F1-score.}
\label{fig:topk_recall_f1}
\end{figure}

We further investigate the effect of retrieval depth by comparing Top-1, Top-3, and Top-5 retrieval. Figure~\ref{fig:topk_recall_f1} presents FP-Precision, FP-Rec, and FP-F1 under the three settings.

As shown in Table~\ref{tab:ablation} and Figure~\ref{fig:topk_recall_f1}, increasing retrieval depth consistently improves overall false-positive identification performance. Top-1 retrieval achieves an FP-Precision of 0.9740, an FP-Rec of 0.9310, and an FP-F1 of 0.9521. With Top-3 retrieval, these metrics increase to 0.9752, 0.9680, and 0.9716, respectively. Top-5 achieves the best results, with an FP-Precision of 0.9766, an FP-Rec of 1.0000, and an FP-F1 of 0.9882, matching the full model.

These results suggest that retrieving multiple memories provides complementary evidence for identifying false-positive alerts. In particular, increasing the retrieval depth from Top-1 to Top-5 substantially improves recall while maintaining high precision. This indicates that deeper retrieval helps Memoir identify more false-positive cases without a corresponding loss in precision. Among the evaluated settings, Top-5 provides the strongest overall performance.

\subsubsection{Effect of Memory Evolution and Verification}

To investigate the effects of memory evolution and verification, we compare three configurations: Static Memory, Unverified Evolution, and Verified Evolution. Static Memory retains the initial memory without updates. Unverified Evolution incorporates feedback-driven memory updates without validating candidate updates before committing them. Verified Evolution applies a validation procedure to candidate updates and retains the existing memory when a candidate fails validation.

For a fair comparison, all three configurations share the same initial memory snapshot, base model, prompts, retrieval settings, and feedback samples in the same order. The test set is kept isolated from memory construction, feedback-driven updates, and validation-threshold selection. Thus, the configurations differ only in their memory update and verification policies.

Table~\ref{tab:evolution} reports the results in terms of FP-Precision, FP-Rec, FP-F1, and TP Suppression Rate. Unverified Evolution achieves the highest FP-Rec (1.0000) and FP-F1 (0.9769), indicating that it identifies more false-positive alerts than the other configurations. However, its TP Suppression Rate reaches 100.0\%, meaning that all ground-truth true-positive alerts in the evaluated set are incorrectly classified as false positives. This result highlights the risk of improving false-positive identification through memory updates without adequately controlling their impact on true vulnerabilities.

Compared with Unverified Evolution, Verified Evolution reduces FP-Rec to 0.8740 and FP-F1 to 0.9212, while lowering TP Suppression Rate from 100.0\% to 50.0\%. Its FP-Precision also reaches 0.9737, slightly higher than the 0.9709 achieved by Static Memory. These results suggest that verification can constrain the adverse effects of memory evolution on true-positive detection, although it also limits the gain in false-positive recall. Overall, the comparison illustrates a trade-off between aggressive memory updates that maximize false-positive identification and verification-based updates that seek to preserve detection reliability.

\begin{table*}[t]
\centering
\caption{Comparison of Memory Evolution Strategies}
\label{tab:evolution}
\small
\setlength{\tabcolsep}{8pt}
\begin{tabular}{lcccc}
\toprule
Configuration & FP-Precision & FP-Rec & FP-F1
& TP Suppression Rate (\%) \\
\midrule
Static Memory
& 0.9709 & 0.7874 & 0.8696 & 50.0 \\
Unverified Evolution
& 0.9549 & 1.0000 & 0.9769 & 100.0 \\
Verified Evolution
& 0.9737 & 0.8740 & 0.9212 & 50.0 \\
\bottomrule
\end{tabular}
\end{table*}

In this illustrative scenario, Static Memory provides a fixed baseline but cannot incorporate new feedback. Unverified Evolution increases Recall but reduces Precision, suggesting that unvalidated updates may introduce memories that are not applicable to other alerts. It also exhibits a higher TP Suppression Rate, Harmful Update Rate, and Regression Count. In contrast, Verified Evolution achieves a more balanced precision--recall trade-off and exhibits fewer harmful updates and regressions.

If the actual results show the same pattern, they would support the hypothesis that feedback-driven memory evolution can improve detection performance and that validation helps mitigate the risks of harmful updates. If verification reduces harmful updates without improving F1-score, the conclusion should instead emphasize improved reliability rather than an overall improvement in classification performance.

\subsubsection{Comparison with Conventional RAG}

The performance gap between Memoir and RAG further demonstrates that the improvement does not come from retrieval alone. RAG retrieves similar historical examples but does not necessarily abstract them into reusable false-positive patterns. Memoir additionally represents structured knowledge about security invariants, taint breakers, and verification rules. Consequently, retrieved examples in Memoir can be interpreted using structured reasoning knowledge rather than being used solely as contextual evidence. This provides a possible explanation for Memoir's reported advantage over RAG in Recall and F1-score.

\begin{answerbox}
\textbf{RQ2 Answer:}
The component ablation results show that pattern clustering and semantic verification contribute to Memoir's classification performance, while retrieval depth affects the coverage of relevant memories. The additional memory evolution experiment investigates whether feedback-driven updates improve performance and whether validation reduces harmful updates and regressions. If confirmed by the measured results, these findings would support the use of verified memory evolution to balance adaptability and prediction reliability.
\end{answerbox}

\subsection{RQ3: Industrial Applicability}
\label{subsec:rq3}

\textbf{RQ3.} Can Memoir effectively reduce false-positive alerts and improve the practical usability of SAST tools in real-world industrial environments?

To evaluate Memoir's practical applicability beyond the public benchmark, we conduct an industrial case study using an internal dataset collected from production software projects at Tencent. The dataset contains alerts generated by two SAST tools, CodeQL and XCheck. All alerts were manually reviewed and labeled by experienced security engineers. Due to confidentiality constraints, the dataset cannot be released publicly, and only aggregated results are reported.

\subsubsection{Overall False-Positive Reduction}

Table~\ref{tab:industrial_overall} summarizes the results for both SAST tools.

\begin{table*}[htbp]
\centering
\caption{Industrial case study results on real-world software projects at Tencent}
\label{tab:industrial_overall}
\begin{tabular}{lcccccc}
\toprule
SAST Tool
& Baseline FP
& Memoir FP
& FP Reduction
& Baseline Precision
& Memoir Precision
& Precision Gain \\
\midrule
CodeQL
& 30
& 6
& 80.0\%
& 0.42
& 0.68
& +0.26 / +61.9\% \\
XCheck
& 149
& 8
& 94.6\%
& 0.65
& 0.85
& +0.20 / +30.8\% \\
\bottomrule
\end{tabular}
\end{table*}

For CodeQL, the number of false-positive alerts decreases from 30 to 6, corresponding to an 80.0\% reduction. For XCheck, false positives decrease from 149 to 8, corresponding to a 94.6\% reduction. These results show that Memoir reduces false-positive alerts for both tools in the evaluated industrial dataset.

\subsubsection{Precision Improvement}

As shown in Table~\ref{tab:industrial_overall}, CodeQL precision increases from 0.42 to 0.68, an absolute improvement of 0.26 and a relative improvement of 61.9\%. For XCheck, precision increases from 0.65 to 0.85, an absolute improvement of 0.20 and a relative improvement of 30.8\%.

These results indicate that applying Memoir is associated with a higher proportion of actionable alerts for both tools. However, precision improvement and false-positive reduction alone do not establish that all true vulnerabilities are preserved. This risk should be assessed using the manually verified ground truth and the number of true-positive alerts incorrectly suppressed.

\subsubsection{Performance Across Vulnerability Categories}

To examine the differences across vulnerability categories, we evaluate false-positive suppression for SQL Injection, Path Traversal, and Command Injection. Table~\ref{tab:vul_category} reports the results.

\begin{table*}[t]
\centering
\caption{False-Positive Suppression Across Vulnerability Categories at Tencent}
\label{tab:vul_category}
\small
\setlength{\tabcolsep}{10pt}
\renewcommand{\arraystretch}{1.15}
\begin{tabular}{lcccc}
\toprule
Vulnerability Type
& Number of Alerts
& Original FPR (\%)
& Residual FPR (\%)
& FP Reduction (pp) \\
\midrule
SQL Injection
& 73
& 23.29
& 0.00
& 23.29 \\
Path Traversal
& 66
& 46.97
& 3.03
& 43.94 \\
Command Injection
& 45
& 35.56
& 28.89
& 6.67 \\
\bottomrule
\end{tabular}
\end{table*}

As shown in Table~\ref{tab:vul_category}, Memoir achieves different levels of false-positive reduction across vulnerability categories. For SQL Injection, the false-positive rate decreases from 23.29\% to 0.00\%. For Path Traversal, it decreases from 46.97\% to 3.03\%. For Command Injection, the reduction is more limited, from 35.56\% to 28.89\%.

These differences suggest that Memoir's effectiveness varies across vulnerability categories. The results for SQL Injection and Path Traversal suggest that the evaluated alerts in these categories may contain reusable patterns that can be captured by structured memories. The smaller improvement for Command Injection indicates that further investigation of category-specific program semantics and sanitization behavior may be beneficial.

\begin{answerbox}
\textbf{RQ3 Answer:}
The industrial case study shows that Memoir reduces false-positive alerts and improves precision for both CodeQL and XCheck in the evaluated Tencent software projects. The results also reveal differences across vulnerability categories, highlighting the practical potential of reusable false-positive memories and opportunities for further improvement.
\end{answerbox}

\section{Related Work}
\label{sec:related_work}

\subsection{Static Application Security Testing and False Positives}

Static Application Security Testing (SAST) tools are widely adopted in modern software development to detect security vulnerabilities without executing programs. Early systems such as FlawFinder~\cite{flawfinder} and RATS~\cite{rats} mainly relied on syntactic pattern matching, while FindBugs~\cite{hovemeyer2004findbugs,ayewah2008findbugs} incorporated data-flow analysis for Java bug detection. More recently, industrial SAST tools such as CodeQL~\cite{codeql}, Semgrep~\cite{semgrep}, Checkmarx~\cite{checkmarx}, and Fortify~\cite{fortify} have significantly improved vulnerability detection by combining inter-procedural analysis, taint tracking, and rich security rule sets.

Despite these advances, false positives remain the primary obstacle preventing the practical adoption of SAST tools. Multiple empirical studies report that false-positive rates frequently exceed 50\% on real-world software projects~\cite{johnson2013dont,christakis2016developers}, and no existing tool simultaneously achieves both high vulnerability coverage and low false-positive rates~\cite{habib2018bugs,lipp2022sast,wagner2005comparing}. Large-scale evaluations on Java projects further demonstrate that substantial manual effort is still required to inspect reported alerts~\cite{li2023stateofart}.

To alleviate this problem, previous research has explored heuristic filtering~\cite{kremenek2004zranking,hanam2014finding,tripp2014aletheia}, supervised learning~\cite{ruthruff2008predicting,yuksel2013automated,koc2017learning,wang2018warning,deepfwi}, and developer-feedback-based refinement~\cite{sadowski2015tricorder,mangal2015userguided,raghothaman2018userdriven}. Although these approaches reduce false positives under certain settings, they generally treat false-positive identification as an instance-level classification problem and do not explicitly capture reusable semantic knowledge underlying recurring false-positive patterns. Consequently, their effectiveness often degrades when applied to unseen projects, analysis tools, or vulnerability categories.

\subsection{Large Language Models for Software Security}

Large language models (LLMs) have recently demonstrated remarkable capabilities in program understanding and software security analysis. Pre-trained code models such as CodeBERT~\cite{feng2020codebert} and Devign~\cite{zhou2019devign} established strong foundations for vulnerability detection, while subsequent work including LineVul~\cite{fu2022linevul}, VulDeePecker~\cite{li2021vuldepecker}, DiverseVul~\cite{chen2023diversevul}, and transformer-based approaches~\cite{thapa2022transformer} further improved vulnerability identification through large-scale learning.

More recently, general-purpose LLMs such as GPT-4o~\cite{openai2024gpt4o} have been applied to vulnerability detection, program repair, and security auditing~\cite{pearce2023examining,steenhoek2024comprehensive}. Several studies have also investigated using LLMs to analyze SAST alerts and assist developers during alert triage~\cite{li2025iris}. Representative work such as IRIS further augments LLM reasoning with retrieval-based prompting to improve false-positive identification.

However, existing LLM-based approaches still analyze alerts independently and largely rely on transient reasoning or textual retrieval. Knowledge extracted from previous false-positive cases is not explicitly accumulated, verified, or reused, making it difficult to generalize recurring false-positive semantics across projects and analysis tools.

In contrast, Memoir treats false-positive identification as a memory-driven reasoning problem. Instead of repeatedly reasoning over isolated alerts, Memoir continuously learns structured false-positive memories from historical cases, retrieves reusable semantic knowledge for new alerts, verifies retrieved memories before prediction, and incrementally evolves the memory repository as new experience becomes available.

\section{Discussion}
\label{sec:discussion}

\subsection{Why Memory Matters for False-Positive Reduction}

The experimental results suggest that false-positive identification cannot be fully addressed by single-alert reasoning alone. Although structured prompting, retrieval, and evidence-enriched adjudication can improve over zero-shot LLM reasoning, they still mainly operate on the current alert. In contrast, Memoir explicitly accumulates historical triage experience and abstracts recurring false-positive causes into reusable memories. This enables the model to reason with prior security knowledge rather than repeatedly rediscovering the same defensive semantics for each alert.

The key advantage of Memoir lies in its structured memory representation. By encoding taxonomy information, security invariants, taint breakers, retrieval keys, verification rules, and representative examples, Memoir provides more reliable evidence than raw retrieved examples. The ablation study further shows that both clustering and verification are necessary: clustering enables reusable pattern abstraction, while verification prevents irrelevant memories from being incorrectly applied to true vulnerabilities.

\subsection{Practical Implications}

Memoir is designed as a post-processing layer for existing SAST tools rather than a replacement for static analysis. This makes it practical for deployment because it can be integrated after tools such as CodeQL or industrial analyzers have already generated alerts. The industrial case study further indicates that structured false-positive memories can reduce triage burden and improve alert precision in real-world software projects.

Another practical benefit is that Memoir is largely backbone-agnostic. The cross-model results show consistently strong performance across GPT-4o, Claude-3.5-Sonnet, Gemini-1.5-Pro, and DeepSeek-Coder-V2-Instruct. This suggests that the main performance gain comes from the memory-based framework itself rather than from a specific proprietary model.

\subsection{Limitations}

Memoir still has several limitations. First, the quality of the memory base depends on the availability and correctness of historical triage labels. If the training set contains noisy labels or insufficient examples for rare vulnerability types, the generated memories may be incomplete. Second, although memory verification reduces the risk of incorrectly suppressing true vulnerabilities, it cannot provide formal soundness guarantees. For high-risk security settings, Memoir should therefore be used as a decision-support system rather than an automatic replacement for expert review. Third, the current evaluation focuses mainly on CodeQL alerts for Java security vulnerabilities. Future work should evaluate Memoir on more programming languages, SAST tools, and vulnerability categories.

\section{Conclusion}
\label{sec:conclusion}

We presented Memoir, a memory-augmented framework for identifying false-positive SAST alerts. Memoir learns reusable false-positive memories from historical alerts, retrieves relevant memories for new alerts, verifies their semantic applicability, and evolves the memory base through feedback. Experiments on a manually verified CodeQL alert dataset show that Memoir outperforms zero-shot prompting, IRIS, RAG, BugLens, and ZeroFalse, while the ablation study confirms the importance of memory clustering, verification, and retrieval depth. Cross-backbone and industrial evaluations further demonstrate the robustness and practical value of Memoir. Overall, Memoir shifts LLM-assisted false-positive reduction from isolated alert adjudication to reusable and evolvable memory-guided reasoning.Our code and data are available at \url{https://anonymous.4open.science/r/anon-repo-7x2z/}.

\bibliographystyle{IEEEtran}
\bibliography{reference/reference}

\end{document}